\documentclass[doublecol, figures]{epl2}
\usepackage{blindtext}
\usepackage[T1]{fontenc}
\usepackage[utf8]{inputenc}
\usepackage{subcaption}
\usepackage{amsfonts, amsmath}
\usepackage{amssymb,mathtools}
\usepackage{amstext}
\usepackage{color} 
\usepackage[table]{xcolor}
\usepackage{graphicx}
\usepackage{parskip}
\usepackage{ragged2e}

\title{Synchronisation zweier Labialpfeifen}
\shorttitle{Synchronisation zweier Labialpfeifen}

\author{Jakub Sawicki\inst{1,2,3,*}}
\shortauthor{J. Sawicki}

\institute{                    
  \inst{1} Potsdam Institute for Climate Impact Research\\
  \inst{2} Berlin University of the Arts \\
  \inst{3} University of Applied Sciences Northwestern Switzerland \\
  * zergon@gmx.net
}

\abstract{Die Synchronisation gekoppelter Orgelpfeifen stellt ein nichtlineares Ph{\"a}nomen dar. Die Untersuchung von Synchronisationseffekten bei Orgelpfeifen anhand eines Modells zeitverzögert gekoppelter Van-der-Pol-Oszillatoren zeigt, dass Synchronisation die Tonhöhenstabilität verbessert, aber auch zu Schalldämpfung führen kann. Die Arnold-Zunge verdeutlicht den Einfluss distanzabhängiger Kopplung auf Stabilität und Frequenzanpassung. Die Ergebnisse stimmen qualitativ gut mit Experimenten überein und betonen die Relevanz nichtlinearer Kopplungsmechanismen in der Orgelakustik.}

\begin{document}
\graphicspath{{figs/}}
\maketitle

\section{Orgelpfeifen und nichtlineare Dynamik}

Die Physik von Orgelpfeifen ist ein interdisziplin{\"a}res Forschungsgebiet, das Konzepte aus der Theorie nichtlinearer dynamischer Systeme \cite{FAB00,BAD13,FLU13}, der aeroakustischen Modellierung \cite{HOW03} und der Synchronisationstheorie \cite{PIK01} integriert. In dieser Studie analysieren wir die Synchronisation benachbarter Orgelpfeifen, ein Ph{\"a}nomen, das sowohl stabilisierende als auch d{\"a}mpfende Effekte auf die Tonh{\"o}he haben kann, insbesondere bei Prospektpfeifen \cite{RAY82,FLE78,STA01}. Aktuelle experimentelle und theoretische Untersuchungen liefern hierzu wichtige Erkenntnisse \cite{ABE06,ABE09,FIS14,FIS16,SAW18a,SAW20,SAW25}. 

Zur Beschreibung der nichtlinearen Wechselwirkungen wird eine einzelne labiale Orgelpfeife als selbstangeregter Oszillator modelliert. Das oszillatorische Element ist das Luftstrahlschwingungssystem, das mit dem Pfeifenk{\"o}rper als Resonator interagiert. Schallwellen, die am Labium entstehen, verst{\"a}rken die Schwingung des Luftblatts. Die f{\"u}r die anhaltenden Schwingungen notwendige Energie stammt aus dem unter der Pfeife befindlichen Druckreservoir. Abel et al. \cite{ABE09} zeigten, dass das Verhalten einer Orgelpfeife durch einen Van-der-Pol-Oszillator beschrieben werden kann. Fischer et al. \cite{FIS14,FIS16} untersuchte die Rolle von Nichtlinearit{\"a}ten in der Schallerzeugung und Synchronisation. Experimentelle Daten zeigen \cite{BER06}, dass komplexe Reflexions- und Interaktionseffekte unbeabsichtigte Schalld{\"a}mpfung bewirken k{\"o}nnen. Dennoch erweist sich das einfache Modell der verz{\"o}gert gekoppelten Van-der-Pol-Oszillatoren als geeignet, um zentrale dynamische Merkmale zu erfassen. Die Ber{\"u}cksichtigung der abstandsabh{\"a}ngigen Kopplung erlaubt eine analytische Untersuchung der Synchronisationsfrequenzen und Bifurkationen an den R{\"a}ndern der Arnold-Zunge, eine mathematische Struktur zur Beschreibung der Synchronisationseigenschaften gekoppelter Oszillatoren. Ein Vergleich theoretischer Vorhersagen mit experimentellen Daten zeigt eine qualitative {\"U}bereinstimmung, insbesondere bez{\"u}glich der nichtmonotonen R{\"a}ndern der Arnold-Zunge, die sich von den linearen Arnold-Zungen in einfacheren gekoppelten Systemen unterscheidet. Zur weitergehenden Analyse nutzen wir die Schnelle Fourier-Transformation (FFT), welche die Rechenkomplexit{\"a}t erheblich reduziert. Dadurch erreichen wir eine hohe {\"U}bereinstimmung zwischen experimentellen und simulierten Ergebnissen.

Zuerst stellen wir einen mathematischen Rahmen basierend auf zwei verz{\"o}gert gekoppelten Van-der-Pol-Oszillatoren als vereinfachtes Modell f{\"u}r gekoppelte Orgelpfeifen vor. Anschließend pr{\"a}sentiert analytische Methoden zur genauen Untersuchung der Synchronisationsph{\"a}nomene, wobei die Arnold-Zunge ein wichtige Rolle spielt. Diese Ergebnisse werden anschließend mit experimentellen Daten verglichen und zusammengefasst.  

\section{Ein Modell f{\"u}r gekoppelte Orgelpfeifen}
\label{sec:model}

Um ein tieferes Verst{\"a}ndnis der Synchronisationsph{\"a}nomene zweier gekoppelter Orgelpfeifen zu erlangen, modellieren wir die Pfeifen durch Van-der-Pol-Oszillatoren mit verz{\"o}gerter Kopplung \cite{SAW18a}:
\begin{equation}
\label{ausgang}
\ddot{x}_i+{\omega_i}^2x_i-{\mu}\left[\dot{x}_i-\dot{f}(x_i)+{\kappa}(\tau)x_j(t-\tau)\right]=0,
\end{equation}
wobei $i,j=1,2$. Diese Gleichungen beschreiben jeweils einen harmonischen Oszillator mit einer intrinsischen Winkelgeschwindigkeit $\omega_i$, erg{\"a}nzt durch eine lineare und nichtlineare D{\"a}mpfung der St{\"a}rke $\mu > 0$. Die nichtlineare D{\"a}mpfung kann durch die Funktion
\begin{eqnarray}
\label{ausgang11}
f(x_i)=\frac{{\gamma}}{3}x_i^3,
\end{eqnarray}
dargestellt werden, wobei $\gamma$ der Anisochronizit{\"a}tsparameter ist und $\dot{f}(x_i)={\gamma} x_i^2 \dot{x_i}$. Die Kopplungsverz{\"o}gerung ist $\tau$, und die verz{\"o}gerungsabh{\"a}ngige Kopplungsst{\"a}rke in Gleichung \eqref{ausgang} wird durch $\kappa(\tau)$ beschrieben. Zur Vereinfachung der Berechnungen nehmen wir im ersten Teil der Arbeit die Kopplungsst{\"a}rke als konstant an ($\kappa(\tau)=\kappa$), jedoch gelten alle analytischen Ergebnisse auch f{\"u}r allgemeine $\kappa(\tau)$. Da f{\"u}r die Synchronisation die Frequenzdifferenz der beiden Oszillatoren von Bedeutung ist, f{\"u}hren wir den Verstimmungsparameter $\Delta \in \mathbb{R}$ ein:

\begin{eqnarray}
\label{ausgang12}
\omega_1^2=\omega_2^2+\mu \Delta. 
\end{eqnarray}

\begin{figure}
\centering
\includegraphics[width=.8\linewidth]{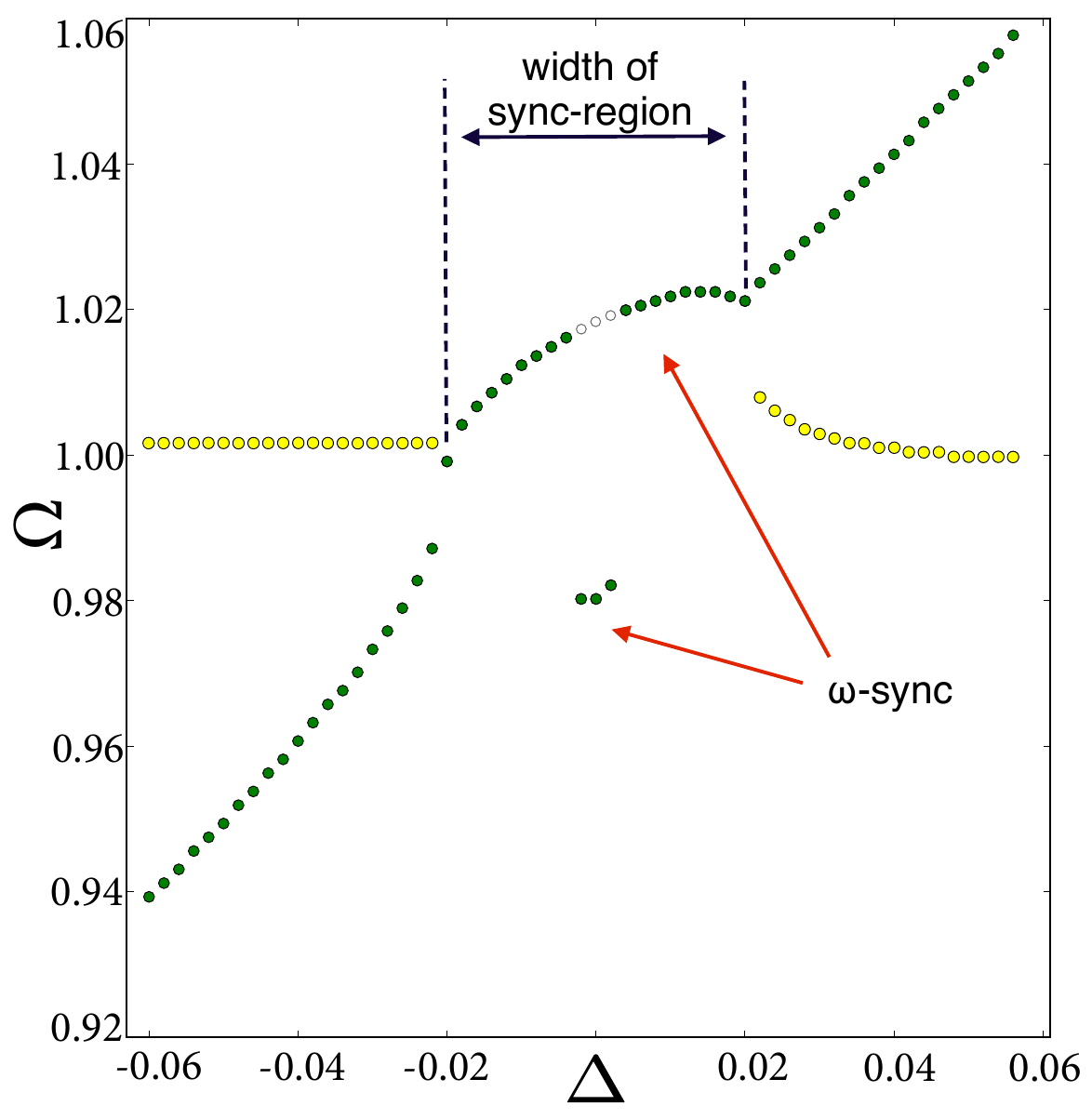}
\caption{\label{goal}\justifying Synchronisation von Orgelpfeifen: Winkelgeschwindigkeit $\Omega$ des Oszillators $x_1$ (dunkelgr{\"u}ne Kreise) und des Oszillators $x_2$ (hellgelbe Kreise) in Abh{\"a}ngigkeit von der Verstimmung $\Delta$ der Oszillatoren. Volle (leere) Kreise entsprechen symmetrischen (nicht-symmetrischen) Anfangsbedingungen. Parameter:  $\omega_{2}=1$, $\mu = 0.1$, $\gamma=1$, $\kappa=0.4$, $\tau=0.1\pi$. Abbildung aus \cite{SAW18a} entnommen.}
\end{figure}

Abbildung~\ref{goal} illustriert das Synchronisationsverhalten, das durch numerische Simulationen der Gleichung \eqref{ausgang} unter symmetrischen Anfangsbedingungen erhalten wurde (linkes Panel). Die Abbildung zeigt die Winkelgeschwindigkeiten $\Omega$ in Abh{\"a}ngigkeit von der Verstimmung $\Delta$ zwischen zwei verz{\"o}gert gekoppelten Van-der-Pol-Oszillatoren. Eine deutliche Synchronisationsregion ist erkennbar, die durch einen abrupten {\"U}bergang gekennzeichnet ist. Innerhalb dieser Region tritt die Gleichphasen synchronisierte L{\"o}sung (unterer Ast der Kurve) nur f{\"u}r kleine Werte von $|\Delta|$ auf, w{\"a}hrend das System f{\"u}r gr{\"o}ßere Verstimmung in den Gegenphasen-synchronisierten Zustand {\"u}bergeht (oberer Ast), selbst wenn symmetrische Anfangsbedingungen gew{\"a}hlt werden (dargestellt durch volle Kreise). Bemerkenswert ist, dass f{\"u}r kleine $|\Delta|$ die Gegenphasen-L{\"o}sung auch auftritt, wenn nicht-symmetrische Anfangsbedingungen angewandt werden (leere Kreise). Ziel dieser Studie ist es, die wesentlichen Merkmale der Synchronisation zu benennen, einschließlich der Synchronisationsfrequenz, der Breite der Synchronisationsregion, der Phasendifferenz im synchronisierten Zustand, ihrer Stabilit{\"a}t und der Bifurkationsszenarien an den Synchronisationsgrenzen.

\section{Die Arnold-Zunge}
\label{sec:analytics}

Die quasiharmonische Reduktion ist eine Methode zur Beschreibung schwach nichtlinearer Schwingungen, bei der die Dynamik über langsam variierende Amplituden und Phasen erfasst wird. Im Fall verschwindender Nichtlinearität reduziert sich das System auf entkoppelte harmonische Oszillatoren mit konstanter Amplitude und Phase. Für kleine Nichtlinearitäten wird angenommen, dass diese Größen zeitabhängig sind, wobei schnell oszillierende und langsam variierende Komponenten getrennt betrachtet werden. Dies ermöglicht eine systematische Näherung über die Methode der Mittelung.

Zur Behandlung von zeitverzögerten Kopplungstermen wird eine Taylor-Entwicklung verwendet, wobei angenommen wird, dass das Produkt aus Kopplungsstärke und Verzögerungszeit klein ist. Diese Vereinfachungen erlauben die Reduktion der ursprünglichen verzögerten Differentialgleichungen auf ein System gewöhnlicher Differentialgleichungen für die Amplituden zweier Oszillatoren sowie für die Phasendifferenz, auch als langsame Phase bezeichnet. Diese Phasengleichung entspricht einer verallgemeinerten Adler-Gleichung, einem etablierten Modell zur Beschreibung der Synchronisation in gekoppelten Oszillatoren.

Die Gleichgewichtslösungen dieser Gleichung stehen für Phasen- und Frequenzsynchronisation, bei der der Phasenunterschied konstant bleibt. Ihre Stabilität kann durch Analyse der Form des nichtlinearen Kopplungsterms beurteilt werden. Dabei zeigt sich, dass eine Gegenphasen-Oszillation in vielen Fällen stabil ist, während die Gleichphasensynchronisation instabil sein kann. Dies steht im Einklang mit experimentellen Beobachtungen, bei denen z.B. die Auslöschung des Schalls bei Phasendifferenzen nahe $\pi$ auftritt.

Zentral für die Beschreibung des Synchronisationsverhaltens ist die sogenannte Arnold-Zunge, die durch die verallgemeinerte Adler-Gleichung quantitativ beschrieben werden kann. Besonders bei verschwindendem Frequenzunterschied (Detuning) lassen sich stabile Synchronisationsmodi identifizieren, die mit typischen experimentellen Mustern, wie der Klangverstärkung oder -auslöschung in Orgelpfeifen, übereinstimmen. In Abb.\,\ref{goal} bleibt der obere Frequenzast im Synchronisationsbereich im stabilen Gegenphasen-Modus f{\"u}r $|\Delta|>0$ (im Einklang mit experimentellen Daten \cite{ABE06}), w{\"a}hrend der untere Ast, d. h. der stabile Gleichphasen-Modus, der sich nahe seinem Instabilit{\"a}tspunkt befindet, nur in einem kleinen Bereich von $\Delta$ beobachtbar ist.

\begin{figure}
\centering
\includegraphics[width=1.\linewidth]{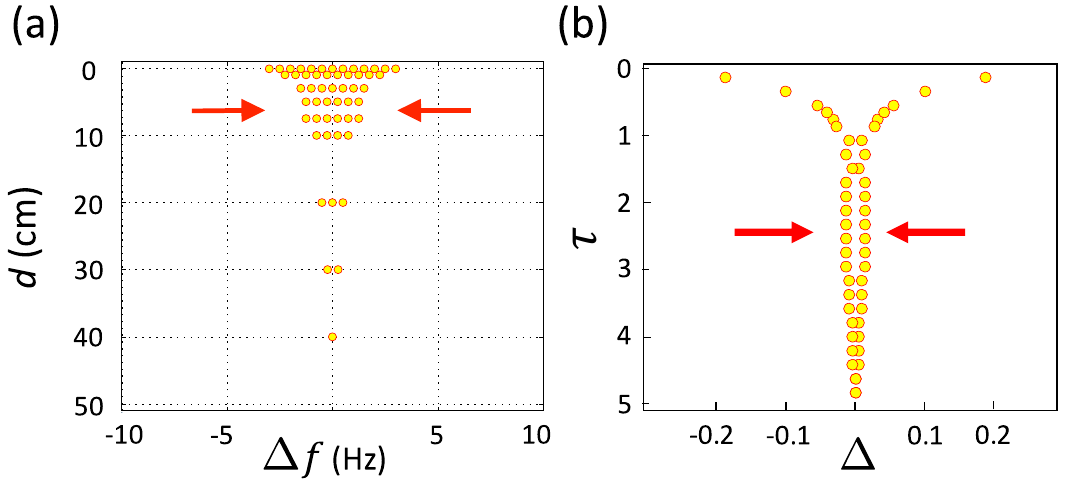}
\caption{\label{coupling_function}\justifying Vergleich von Experiment und Theorie: Arnold-Zunge im Raum der Kopplungsst{\"a}rke $\kappa$ vs. Detuning $\Delta$: (a) Experiment \cite{FIS14,FIS16}, bei dem die Kopplungsst{\"a}rke durch die Distanz $d$ (cm) zwischen den Orgelpfeifen gegeben ist, und (b) numerisches Ergebnis von Gleichung~\eqref{ausgang} mit $\omega_2=1$, $\mu = 0.1$, $\gamma=1$, $\kappa_1=\kappa_n=\kappa_2=\kappa_f=0.04$. Die roten Pfeile in (a) und (b) weisen auf das nicht-monotone Verhalten der Arnold-Zunge hin.}
\end{figure}

Um das experimentell beobachtete, nicht-monotone Verhalten der Arnold-Zunge genau zu erfassen, insbesondere bei kleinen Verz{\"o}gerungszeiten $\tau$ (siehe Abb.\,\ref{coupling_function}a f{\"u}r $d=5\,\text{cm}$), ist es notwendig eine zeitverz{\"o}gerungsabh{\"a}ngige Kopplungsst{\"a}rke $C(\tau)$ einzuf{\"u}hren. Dieses verfeinerte Modell ber{\"u}cksichtigt sowohl Nahfeld- ($\sim \tfrac{1}{\tau^2}$) als auch Fernfeldbeitr{\"a}ge ($\sim \tfrac{1}{\tau}$), indem anstelle eines konstanten Kopplungskoeffizienten $\kappa$ in Gleichung~\eqref{ausgang} eine Kopplungsst{\"a}rke $\kappa(\tau)$, die explizit von der Verz{\"o}gerungszeit $\tau$ abh{\"a}ngt, verwendet wird:

\begin{equation}
\label{eq:near_far}
\begin{split}
\kappa(\tau) = \frac{\kappa_1}{\tau^2} + \frac{\kappa_2}{\tau},
\end{split}
\end{equation}
wobei $\kappa_1=\kappa_n$ und $\kappa_2=\kappa_f$ die Nahfeld- bzw. Fernfeld-Kopplungsparameter sind und $\tau$ die Verz{\"o}gerungszeit darstellt. Eine wesentliche Folge der verz{\"o}gerungsabh{\"a}ngigen Kopplungsst{\"a}rke ist die Destabilisierung der Gleichphasen-L{\"o}sung f{\"u}r $\tau=0$ \cite{SAW18a}, so dass f{\"u}r $\tau=0$ nur der Antiphase-Modus stabil ist, was zu einer Abschw{\"a}chung des Pfeifenklanges führt.

\section{Vergleich mit Experimenten} 
\label{sec:comparison}

Ein direkter Vergleich zwischen einem physikalisch komplexen Experiment und einem vereinfachten Oszillatormodell wirft zentrale Fragen hinsichtlich der Modellierbarkeit akustischer Systeme auf. Orgelpfeifen zeichnen sich durch ein reiches Obertonspektrum, nichtlineare aeroakustische Wechselwirkungen und Phänomene wie Turbulenzen, Grenzschichteffekte und Rückkopplung aus. Demgegenüber abstrahieren mathematische Modelle stark, um wesentliche Dynamiken erfassbar zu machen. Ein prominentes Beispiel ist der Van-der-Pol-Oszillator, der trotz seiner Einfachheit zentrale Eigenschaften selbst-erhaltener Schwingungen wie Grenzzyklen, Bifurkationen und Synchronisation abbildet. Der Wert solcher Modelle liegt in ihrer Fähigkeit, fundamentale Prinzipien zu erfassen. In der Theorie dynamischer Systeme gilt es als etabliert, dass komplexes Verhalten aus einfachen Modellen hervorgehen kann. Unsere Analyse zeigt, dass zentrale Synchronisationsphänomene in Orgelpfeifen qualitativ und quantitativ mit dem Van-der-Pol-Modell reproduzierbar sind. Dies schafft ein analytisches Rahmenwerk, das sich auch auf andere gekoppelte Oszillatorsysteme -- etwa in der Biologie, Mechanik oder Elektronik -- übertragen lässt.

Zur Analyse des Frequenzverhaltens numerischer und experimenteller Daten verwenden wir die schnelle Fourier-Transformation (FFT). Als effizienter Algorithmus zur Berechnung der diskreten Fourier-Transformation ($O(N \log N)$ gegenüber $O(N^2)$) ist sie essenziell für die spektrale Analyse in Echtzeitanwendungen. In unserer Studie dient die FFT der Extraktion dominanter Frequenzen aus Messdaten und Simulationen, was den Vergleich der spektralen Eigenschaften sowie die Identifikation von Synchronisationsregimen und Übergängen ermöglicht. Darüber hinaus erlaubt die Analyse höherer Harmonischer Rückschlüsse auf nichtlineare Effekte und Kopplungsmechanismen.

\begin{figure}
\centering
\includegraphics[width=1.\linewidth]{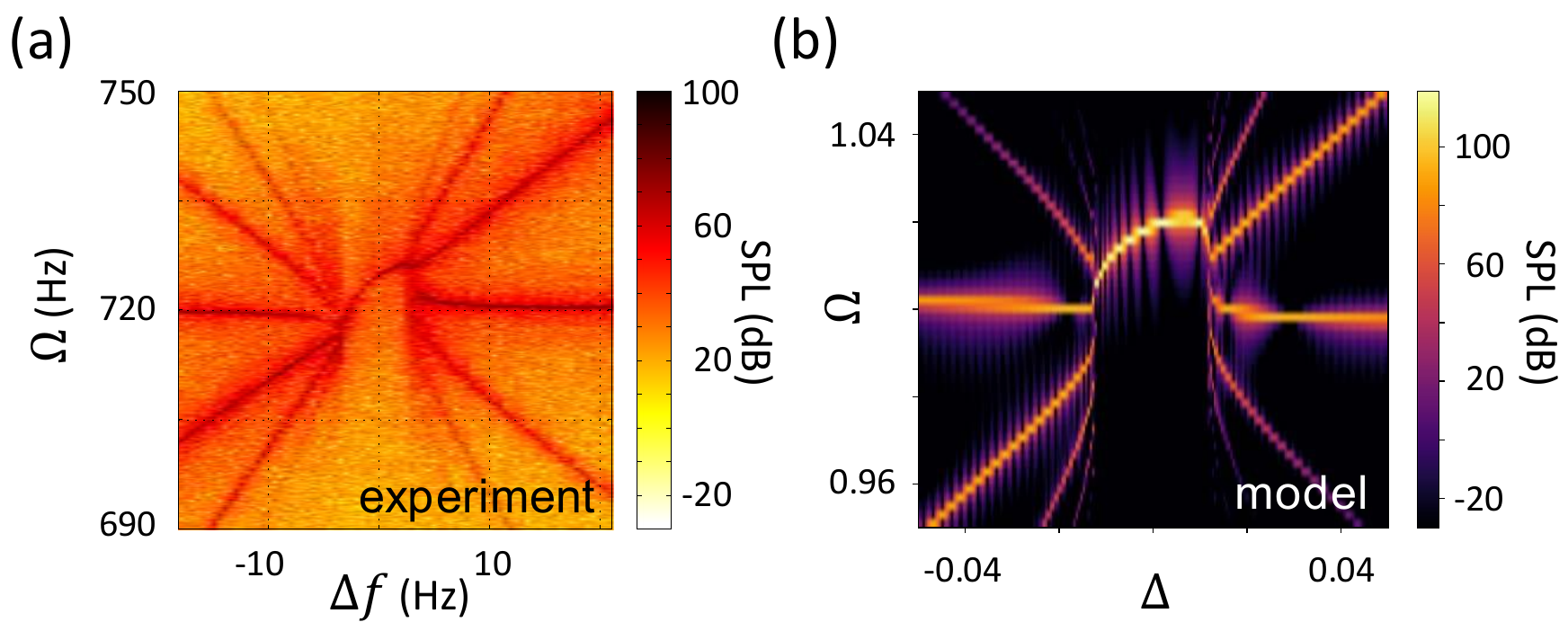}
\caption{\label{realpipeex}\justifying Vergleich der Synchronisationsregionen in Experiment und Theorie: (a) experimentell beobachteter Schalldruckpegel (SPL) im Raum der beobachteten Frequenz $\Delta$ versus Frequenzdetuning $\Delta f$ (in Hz) \cite{FIS14}, und (b) numerisch berechneter SPL unter Verwendung der schnellen Fourier-Transformation (FFT) im Raum der Winkelgeschwindigkeit $\Omega$ versus dimensionsloses Detuning $\Delta$, mit den Parametern $\omega_2=1$, $\mu = 0.1$, $\gamma=1$, $\kappa=0.4$, $\tau=1.1\pi$.}
\end{figure}

Die Synchronisationsszenarien aus unserem Modell stimmen qualitativ gut mit den experimentellen Beobachtungen überein, insbesondere hinsichtlich der Struktur der Übergangsregionen und der charakteristischen Krümmung der Synchronisationsbereiche in Abb.\,\ref{realpipeex}. Dies bestätigt die Relevanz niedrigdimensionaler Modelle zur Beschreibung der zugrunde liegenden Dynamik und stärkt deren Aussagekraft für reale nichtlineare Systeme. Ein weiterer zentraler Aspekt ist die Multistabilität: Verschiedene Anfangsbedingungen können bei gleichen Parametern zu unterschiedlichen Synchronisationszuständen führen. Dies wurde numerisch beobachtet und ist typisch für nichtlineare Systeme. Im Kontext gekoppelter Orgelpfeifen bedeutet dies, dass kleine Änderungen, etwa in der Phasendifferenz oder im Aufbau, das Synchronisationsverhalten beeinflussen können.

In Abb.\,\ref{coupling_function} vergleichen wir experimentelle und analytisch berechnete Arnold-Zungen im Raum der Kopplungsstärke $\kappa$ und des Detunings $\Delta$. Diese keilförmige Synchronisationsregion hängt wesentlich vom Zusammenspiel beider Parameter ab. In Experimenten wird $\kappa$ durch den Abstand der Pfeifen bestimmt, der die akustische Kopplung reguliert. Unsere Ergebnisse zeigen Arnold-Zungen mit deutlich gekrümmten Rändern auch bei kleinen Verzögerungen $\tau$, was die Rolle nichtlinearer Kopplungseffekte unterstreicht \cite{FIS14,FIS16}. Die Einführung einer verzögerungsabhängigen Kopplungsfunktion $\kappa(\tau)$ ermöglicht eine verbesserte Modellierung dieser Effekte und liefert tiefere Einsichten in die zeitverzögerte Synchronisationsdynamik.


\section{Schlussfolgerungen} 
\label{sec:conclusion}

Die Studie analysiert die Synchronisation von Orgelpfeifen unter Berücksichtigung zeitverzögerter Kopplung. Durch Modellierung mit zwei Van-der-Pol-Oszillatoren, die über dissipative und verzögerte Kopplung interagieren, wird das Zusammenspiel von Nah- und Fernfeldmechanismen untersucht. Es zeigt sich, dass die Kopplungsverzögerung -- bedingt durch den Abstand zwischen den Pfeifen -- entscheidend für die Entstehung und Stabilität synchroner Zustände ist.

Analytisch wird das System durch Mittelwertbildung und eine verallgemeinerte Adler-Gleichung beschrieben, ergänzt durch die Beschreibungsfunktionsmethode zur Bestimmung der Synchronisationsfrequenz. Die resultierenden Synchronisationsmuster, insbesondere die Struktur der Arnold-Zunge, spiegeln die komplexe Abhängigkeit von Kopplungsstärke und Verzögerung wider. Dabei treten Gleichphasen- und Gegenphasenmodi auf, deren Stabilität durch die Verzögerung stark beeinflusst wird.

Die Ergebnisse verdeutlichen die Notwendigkeit, sowohl Nah- als auch Fernfeldkopplung in Modellen akustischer Oszillatorsysteme zu berücksichtigen. Darüber hinaus liefern sie allgemeine Erkenntnisse zur Synchronisation in verzögert gekoppelten Systemen mit Relevanz für physikalische, technische und akustische Anwendungen.

\section{Danksagung}
Dank geht an Markus Abel, Jost Fischer, Markus Radke, Eckehard Sch{\"o}ll und Natalia Spitha. Besonderer Dank gilt dem Hausorgel-Arbeitskreis der Gesellschaft der Orgelfreunde e. V., dessen Jahrestreffen 2024 ein Forum f{\"u}r die Diskussionen {\"u}ber die hier zusammengefassten Ergebnisse bot.

\end{document}